\documentclass[12pt,preprint]{aastex}

\shorttitle{Diffuse Gas near IC~443}
\shortauthors{Hirschauer et al.}

\begin{document}
\title{Diffuse Atomic and Molecular Gas near IC~443}
\author{A. Hirschauer\altaffilmark{1}, S. R. Federman\altaffilmark{2}, 
George Wallerstein\altaffilmark{3}, and T. Means\altaffilmark{3}} 
\altaffiltext{1}{Department of Physics, University of Notre Dame, Notre 
Dame, IN 46556.}
\altaffiltext{2}{Department of Physics and Astronomy, University of Toledo, 
Toledo, OH 43606; steven.federman@utoledo.edu}
\altaffiltext{3}{Department of Astronomy, University of Washington, 
Seattle, WA 98195; wall@astro.washington.edu}

\begin{abstract}

We present an analysis of results on absorption from Ca~{\small II}, 
Ca~{\small I}, K~{\small I}, and the molecules CH$^+$, CH, C$_2$, 
and CN that probes gas interacting with the supernova remnant 
IC~443.  The eleven directions sample material across the visible 
nebula and beyond its eastern edge.  Most of the neutral material, 
including the diatomic molecules, is associated with the ambient 
cloud detected via H~{\small I} and CO emission.  Analysis of 
excitation and chemistry yields gas densities that are typical of 
diffuse molecular gas.  The low density gas probed by Ca~{\small II} 
extends over a large range in velocities, from $-$120 to +80 km 
s$^{-1}$ in the most extreme cases.  This gas is distributed among 
several velocity components, unlike the situation for the shocked molecular 
clumps, whose emission occurs over much the same range but as very broad 
features.  The extent of the high-velocity absorption suggests a shock 
velocity of 100 km s$^{-1}$ for the expanding nebula.

\end{abstract}


\section{Introduction}

IC~443 seems to be an example of the interaction of a supernova remnant
(SNR) with an interstellar cloud or complex of clouds.  As such it provides 
the opportunity to observe the effects of the SN shock penetrating the high
density gas of the clouds.  This involves heating the cloud to x-ray
temperatures, followed by rapid cooling leading to a significant increase
in the cloud's density.  IC~443 has been a very popular target and
comparison object primarily for x-ray observations and as an object with
emission by excited molecules.  According to SIMBAD there have been 619
published studies or references to published studies between 1850 and
2007.

IC~443 has provided an opportunity to compare observations and theory of
the hydrodynamics and radiative transfer associated with the penetration
of clouds by a SN shock.  Specific calculations were first made by McKee
\& Cowie (1975) and reviewed by Chevalier (1977).  Early detailed
spectroscopic observations of absorption from the SN shocked clouds 
associated with the Vela Remnant were described by Jenkins, Silk, \& 
Wallerstein (1976).  Here we describe measurements on atomic and molecular 
absorption seen in the spectra of stars behind IC~443 and attempt to 
link them to the wealth of results already available.

Observations of SNRs may be made by a wide variety of instruments each of
which defines the spectral resolution, spectral coverage, and spatial
resolution.  We present new data on interstellar (IS) absorption 
from atoms and molecules at visible wavelengths.  Such observations provide 
the finest spatial resolution and high spectral resolution, but are 
limited by the availabilty of background stars.  The stars must be of 
sufficient brightness and have a relatively simple absorption spectrum to 
permit a clear definition of the IS features.  Members of the Gem OB1 
association afford us the opportunity to probe significant portions of the 
sky in the vicinity of IC~443.

Early molecular observations with good spatial resolution of phenomena 
associated with IC~443 were obtained by Cornett, Chen, \& Knapp (1977) and 
Scoville et al. (1977) using the CO line at 2.6 mm.  The former achieved a 
resolution of 24$^{\prime\prime}$ $\times$ 27$^{\prime\prime}$ while the 
latter had a half-power beamwidth of 2$^{\prime}$.  Both
studies compared their contours with the red image of the Palomar Sky
Survey where the emission is dominated by H-alpha.  The strongest emission
was observed to come from the region between the two most prominent
H-alpha emitting clouds, hinting that the CO was associated with a molecular 
cloud rather than with the nearby H~{\small II}
regions.  In fact most papers since then have referred to IC~443 as the
interaction of a SNR with a molecular cloud.  As such it is probably the
most readily observable example of an SNR associated with a molecular
cloud because both the Vela Remnant and the Cygnus Loop seem to involve
atomic gas with only a small amount of associated molecules.

Two papers within the past few years have attracted attenion to IC~443 once
again.  Snell et al. (2005) used the {\it Submillimeter Wave Astronomy 
Satellite} to observe previously inaccessable lines of H$_2$O, O$_2$, 
C~{\small I}, and $^{13}$CO along with ground-based observations of CO and 
HCO$^+$.  The authors invoke a combination of a fast J-type shock with 
velocity of about 100 km s$^{-1}$ and a slow shock (10 to 20 km s$^{-1}$) 
that could be either J-type of C-type.  Using the Very Large Array and Arecibo 
telescopes Lee et al. (2008) observed 21 cm H~{\small I} emission with a 
resolution of 40 arcseconds, comparable to that of the CO emission of 
Cornett et al. H~{\small I} emission with a velocity range from $-$100 to
$+$50 km s$^{-1}$ was found.  The systemic velocity was $-$5 km s$^{-1}$, 
like that found for the ambient molecular cloud.   The high spectral 
resolution allowed Lee et al. to resolve the shocked H~{\small I} 
into filaments, and they proposed that the main shock propagated
through a uniform medium before reaching the edge of the cloud and freely 
expanding toward the southeast.  Much of the H~{\small I} emission in the 
northeast portion was attributed to recombining gas behind the shock 
front, while the southern emission was coincident with the dissociative 
molecular shock.

Our main goal is to examine how results from atomic and molecular absorption 
fit into the above picture.  We also build upon the earlier measurements 
on absorption from Na~{\small I} D and Ca~{\small II} K by Welsh \& 
Sallmen (2003), who studied four stars at relatively high spectral 
resolution.  They detected absorption from numerous components over a 
velocity range from $-$100 to $+$50 km s$^{-1}$, much like that seen in 
H~{\small I} emission.  Our sample enlarges the number of sight lines probed 
and incorporates measurements on Ca~{\small I}, K~{\small I}, and the 
molecular species CH$^+$, CH, C$_2$, and CN.  These species help us make 
connections between the diffuse molecular gas and the ambient molecular 
cloud.  The result is a more complete understanding of the interaction 
between a SN shock and a molecular cloud.

\section{Observations}

Observations were obtained with the echelle spectrograph of the 3.5-m
telescope of the Apache Point Observatory on 21 March and 28 December of 
2006, 5 and 28 January of 2007, and 3 January of 2008.  Our targets are 
listed in Table 1, along with pertinent stellar data.  The 
coordinates and spectral type, as well as the $V$ and $B-V$ magnitudes, 
are from the SIMBAD database, operated by CDS, Strasbourg, France.  The 
absolute magnitude and reddening are derived from these data 
(see B\"{o}hm-Vitense 1981).  Figure 1 
displays the position of the lines of sight relative to CO contours 
of Huang, Dickman, \& Snell (1986).  The extent of the radio continuum 
emission is indicated by the dashed line.  Bias frames, 
flat fields, and Th-Ar comparison spectra were acquired each night.  
The bias subtracted, flat fielded, wavelength calibrated, 
one-dimensional spectra were extracted with IRAF routines.  Doppler 
corrections were applied to bring the spectra into the frame of the 
Local Standard of Rest.  All spectra have a resolving power of about 
35,000 and a signal-to-noise ratio from about 200 at 8000 \AA\ declining 
slowly to between 50 and 100 at the CN lines near 3875 \AA.  
Representative spectra appear in Figures 2 to 4.

The moderate-resolution spectra offered a glimpse at the component 
structure for the directions probing the vicinity of IC~443.  Using 
standard routines in the IRAF environment, we extracted equivalent 
widths ($W_{\lambda}$) in the following way.  We started the process 
by obtaining the structure seen in the K~{\small I} $\lambda$7699 
spectra; the features are relatively strong and many of the 
components are present.  We then found the amount of absorption 
from the species showing much weaker absorption (Ca~{\small I} and 
the molecules CH, CH$^+$, CN, and C$_2$), making certain of the 
reality of these components by requiring consistency in $v_{LSR}$ 
among the species in light of the spectral resolution.  The 
high-resolution study by Pan et al. (2004) helped us determine the 
expected correspondences.  The last step involved the extraction of 
the Ca~{\small II} components.  The availability of lines from both 
Ca~{\small II} H and K aided the process, especially when the number 
of components was much greater than seen in other species.  The 
Ca~{\small II} absorption toward HD~43582 was particularly 
challenging; in this case we were guided by the results from Welsh 
\& Sallmen (2003), but we note that our final component structure 
differed somewhat from theirs.  Our survey does not incorporate results 
on the Na~{\small I} D lines; the substantial optical depth at line 
center and our coarse spectral resolution did not allow 
meaningful interpretation.  Tables 2 and 3 present the 
compilation of our set of $W_{\lambda}$s.

Special care is required when deriving column densities from spectra 
that may contain unresolved component structure.  We applied several 
methods to minimize the potential impact of unresolved structure.  
First, the `effective' $b$-value for the dominant, molecule-bearing 
components was determined by requiring the absorption from K~{\small I} 
$\lambda$7699 and from the much weaker doublet at 4044 and 4047 
\AA, which was seen toward HD~254755, to yield the same column 
density.  Because many of the lines are not 
optically thin, this was necessary to account for optical depth effects 
in curves of growth.  A self-consistent column density indicated an 
`effective' $b$-value of 2.2 km s$^{-1}$.  The detailed analysis of 
component structure in Ca~{\small II}, Ca~{\small I}, K~{\small I}, 
CH$^+$, CH, and CN by Pan et al. (2005) revealed a typical 
$b$-value for K~{\small I} lines of 1 km s$^{-1}$, suggesting that 
unresolved structure is indeed present in our spectra.  It appears 
that there are two components separated by about 2 km s$^{-1}$ 
hidden within the features seen in our data.  According to 
Pan et al. (2005), K~{\small I}, Ca~{\small I}, CH, and CN lines 
have comparable $b$-values, and we adopted a $b$-value of 2.2 km s$^{-1}$ 
for the latter three species in an analysis based on curves of 
growth.  For CH$^+$, whose typical $b$-value is somewhat larger 
(Pan et al. 2005), we chose $b$ $=$ 2.5 km s$^{-1}$.  The suite of 
molecular oscillator strengths used here comes from (Federman et al. 
1994) and the atomic $f$-values are from Morton (1991), for consistency 
with our earlier work.  The differences with Morton (2003) are only at the 
level of a few percent.

Second, the complete spectra afforded by the echelle spectrograph 
allowed us to compare results from strong and weak lines from CH$^+$, 
CH, C$_2$, and CN.  In particular, there are (1) CH$^+$ 
$\lambda\lambda$3957,4232, (2) the $A-X$ and $B-X$ transitions in CH, 
(3) P, Q, and R lines for C$_2$, and (4) the P and R lines of CN.  
Furthermore for CH, the lines at 3878 and 3890 \AA\ arise from the 
same $\Lambda$-doubling component of the ground state, and earlier 
studies at visible wavelengths (Danks, Federman, \& Lambert 1984; Lien 
1984; Jura \& Meyer 1985) indicate that the populations for both 
$\Lambda$-doubling components are the same.  Thus, the column density 
obtained from $\lambda$3886 should equal that from the two lines of the 
other $\Lambda$-doubling component.  Within the mutual uncertainties of 
the features detected in our spectra, self-consistent column densities 
are obtained in all cases.  In the cases of C$_2$ and CN, where 
absorption from multiple rotational levels is seen, total column 
densities are also obtained for the chemical analyses described below.  
The summary of results for C$_2$ is given in Table 4.  When only $J$ 
$=$ 4 shows detectable absorption, or when absorption is only seen for 
$J$ $=$ 0 to 4, the total C$_2$ column density is estimated from 
the distribution of levels seen toward HD~254577.  In the former case, 
the result for $J$ $=$ 4 is multiplied by 3.3, while the sum for $J$ $=$ 
0 to 4 is multiplied by 1.5 in the latter.  When comparison is made 
with high-quality results based on the $D-X$ (0,0) band at 2313 \AA\ 
(Lambert, Sheffer, \& Federman 1995; Sonnentrucker et al. 2007), the 
multiplicative factors yield total column densities that are 25-40\% 
smaller.  However, most of the difference arises from levels with 
$J$ $>$ 10.  When only the $J$ $=$ 0 line of CN is detected, 
the CN column density is inferred by multiplying 
the result for the ground state by 1.5, a result consistent with 
excitation from the 2.7 K Cosmic Background.

For Ca~{\small II} H and K, we adopted the Doublet Ratio Method 
(Str\"{o}mgren 1948; M\"{u}nch 1968).  A listing of the set of 
column densities for each component seen in Ca~{\small II} absorption 
appears in Table 5; the velocities come from the results for 
Ca~{\small II} K.  A comparison of the Ca~{\small II} column densities 
toward HD~254577, HD~43582, and HD~254755 found by Welsh \& Sallmen 
(2003) generally shows reasonable ageement, with the most optically 
thick velocity components agreeing to better than 50\%.  With our 
limited spectral resolution and their limited signal-noise, better 
correspondence cannot be expected.

\section{Results and Discussion}
\subsection{General Results}

Before describing specifics, we note some general findings.  The most 
widespread absorption is found on the eastern portion of the remnant.  
The Ca~{\small II} spectra seen toward HD~254477, HD~254577, and 
HD~43582 reveal components spanning the range from $-$120 to $+$80 
km s$^{-1}$ (in the case for HD~43582).  Moreover, Ca~{\small II} features 
are present at $-$75 km s$^{-1}$ toward HD~254346, HD~254755, and HD~43907; 
these directions are relatively close to the three sight lines with the 
widest ranges in velocity components (see Fig. 1).  The remaining 
directions in our survey probe more northern material.  
While absorption from the neutrals, Ca~{\small I} and 
K~{\small I}, is more restricted, there are components at $-$80 km 
s$^{-1}$.  On the other hand, most of the molecular absorption occurs 
at low velocities (between $-$10 and $+$10 km s$^{-1}$).  The main 
components at slightly negative velocities correspond to the 
ambient gas detected via H~{\small I} (Lee et al. 2008) and CO 
emission (Cornett et al. 1977; Scoville et al. 1977) and to the peaks 
seen in CO emission from the preshock molecular clumps (e.g., Huang et al. 
1986; Dickman et al. 1992).  CH$^+$ is also detected at intermediate 
velocities ($-$25 to $-$75 km s$^{-1}$) toward HD~254477, HD~254577, 
and HD~43703, which comes as a surprise.

General trends are also seen in abundances.  Some of the Ca~{\small I} 
lines are relatively strong with $W_{\lambda}$ $\sim$ 10 m\AA\ and 
in the case of the $-$1.2 km s$^{-1}$ component toward HD~254477, it 
is very strong ($W_{\lambda}$ $\sim$ 50 m\AA).  The ratio of Ca~{\small I} 
to Ca~{\small II} column densities, however, is within a factor of a few 
of 0.01, which is typical of the general ISM (Welty, Hobbs, \& Morton 
2003).  This atomic material probably has gas densities of about 10 
cm$^{-3}$, as suggested by the analysis of H~{\small I} emission (Lee et 
al. 2008).  Significant amounts of molecular absorption are seen in many 
directions, indicating that diffuse molecular gas lies beyond the CO contours 
in the map of Huang et al. (1986) and even the contours of the ambient cloud 
(Cornett et al. 1977; Scoville et al. 1977).  The most molecular rich 
diffuse gas, as revealed by CN, appears to be associated with clumps 
B, C, and D (Huang et al. 1986).  HD~43582 must lie in front 
of clump H because C$_2$ and CN are not detected.  Finally, we note that 
HD~43871 lies in front of IC~443; this sight line contains but small 
amounts of Ca~{\small II} and K~{\small I}, a result consistent with the 
much lower reddening toward the star.

\subsection{Molecular Analysis}

Our spectra allow us to derive the physical conditions, gas density and 
temperature and the flux of ultraviolet (UV) radiation permeating the 
cloud(s), from the distribution of C$_2$ rotational levels and from the 
chemistry involving CH, C$_2$, and CN.  We adopted the method of 
Federman et al. (1994), with updates given in our more recent papers 
(e.g., Pan et al. 2005).  The discussion in Sheffer et al. (2008) 
provides a sense of the limitations associated with the chemical analysis.  
Since we are seeking densities good to a factor of two, these limitations 
are less of a concern here.

\subsubsection{C$_2$ Excitation}

The excitation of C$_2$ molecules arises from a combination of collisions, 
which mainly affects the populations of low-lying rotational levels, and 
pumping via infrared photons to excited electronic states followed by 
cascades (van Dishoeck \& Black 1982).  Our focus is on determinations of 
gas density and temperature.  We compared in a least-squares 
fashion the observed rotational populations relative to the population 
in $J$ $=$ 2 with the predicted ratios from van Dishoeck (1984), whose 
tables of results are more extensive than those of van Dishoeck \& Black 
(1982).  The outcome is the parameter, 
$n$ $\sigma_o$ / $I_{ir}$, for a given kinetic temperature.  Here, the 
density of collision partners, $n$, equals $n$(H) $+$ $n$(H$_2$), 
$\sigma_o$ is the cross section for collisional de-excitation, and $I_{ir}$ 
is the enhancement in the infrared flux relative to the average 
interstellar value.  We continue to use a cross section of $2 \times 10^{-16}$ 
cm$^2$ (van Dishoeck \& Black 1982), but note that the calculations of
Lavendy et al. (1991), Robbe et al. (1992), and Najar et al. (2008)
indicate values a factor of two larger.  We also assume that there is no
enhancement in the infrared radiation field ($I_{ir}$ $=$ 1).  Since we are 
interested in the gas density, $n_H$ $=$ $n$(H) $+$ 2 $n$(H$_2$), we 
multiply $n$ by 1.5, as we have done in the past.  Finally, we have to take 
into account the fact that we used slightly smaller oscillator strengths 
for the $A-X$ electronic transitions.  [Our band oscillator strength of 
$1 \times 10^{-3}$ is within 1-$\sigma$ of the recommended revised value 
of Lambert et al. (1995).]  Our densities are divided by 
1.6 to account for this difference.  The analysis provides values for gas 
densities that are known to a factor of two and for kinetic temperatures 
that are consistent with the relative populations in the lowest 
rotational levels.

We obtained results for several rotational levels toward ALS~8828, 
HD~254346, \linebreak 
HD~254577, and HD~254755.  However, the results for ALS~8828 
were not especially meaningful.  For the other three sight lines, we 
infer kinetic temperatures and gas densities of 30 to 60 K and 200 to 
450 cm$^{-3}$, 10 to 60 K and 200 to 450 cm$^{-3}$, and 30 to 40 K and 
200 cm$^{-3}$ for the diffuse molecular gas toward HD~254346, HD~254577, 
and HD~254755, respectively.  Such values are typically found for diffuse 
material, but the densities are significantly lower than the values 
derived for the preshock molecular gas, 3000 to 10$^4$ cm$^{-3}$, by 
Ziurys, Snell, \& Dickman (1989) and van Dishoeck, Jansen, \& 
Phillips (1993) from molecular emission.  This is not unexpected 
considering the critical densities needed for molecular emission.

\subsubsection{Chemistry}

This analysis is based on a set of algebraic expressions describing the 
rate equations for C$_2$ and CN (e.g., Federman et al. 1994; Pan et al. 
2005; Sheffer et al. 2008).  Here the main outcome is gas density known 
again to a factor of 2.  The reaction rate coefficients, rates for 
photoprocessing, and the fractional abundances for C$^+$, N, and O are 
from our recent work.  A kinetic temperature of 60 K is used, but the 
value does not greatly affect the outcome for gas density.  The flux of 
UV radiation impinging on the cloud is taken to be the average 
interstellar value (i.e., $I_{uv}$ $=$ 1).  This leaves 
the optical depth at 1000 \AA\ for grain attenuation, $\tau_{uv}$, as the 
one significant unknown parameter for each line of sight.  The UV 
extinction seen toward HD~43818, another member of the Gem OB1 
association, by Savage et al. (1985) and Valencic, Clayton, \& Gordon 
(2004) indicates that a typical Galactic extinction law applies.  Thus, 
we used $\tau_{uv}$ $=$ 2 $\times$ 3.1 $\times$ $E$($B-V$) in our 
calculations.  The observed column densities of CH and CN, as well as 
C$_2$ when available, are input for the algebriac expressions, and a 
least-squares analysis that matched observed and predicted CN 
and C$_2$ column density yields gas density.  Before doing so however, two 
additional issues had to be addressed.

Since the stars are about 1500 pc away, we also had to estimate the 
amount of extinction from foreground material not participating in the 
interaction with IC~443.  We considered the following points.  First, 
most of the CH and all of the C$_2$ and CN absorption arises from the 
ambient cloud at velocities near $-$5 km s$^{-1}$.  
This cloud also has at least 85\% of the K~{\small I} 
column along the lines of sight.  Second, we obtained the amount of 
Na~{\small I} D not at the velocity of the ambient molecular gas from the 
measurements of Welsh \& Sallmen (2003) for HD~43582.  This sight 
line, without detectable amounts of CN and $E$($B-V$) of about 0.60, 
contains about half the Na~{\small I} in the molecular velocity component.  
We then suggest that a foreground contribution of $E$($B-V$) $=$ 0.30 
($\tau_{uv}$ $=$ 1.86) applies to the CN-rich directions in our 
sample.

Finally, we took into account that some of the CH was associated with 
CH$^+$ synthesis in low density gas ($n_H$ $<$ 100 cm$^{-3}$), 
not the chemistry of CN (e.g., Lambert, Sheffer, \& 
Crane 1990; Pan et al. 2005).  The components revealing only absorption 
from CH and CH$^+$ were used to correct for this chemical route.  For 
these components, the average $N$(CH$^+$)/$N$(CH) ratio was $0.43 \pm 0.03$.  
For the molecular-rich components at $-$5 km s$^{-1}$, we reduced the 
CH column density by the CH associated with CH$^+$ using the above 
ratio.  The procedure lowered $N$(CH) by 20 to 50\%.

In the course of performing this analysis, we were not able to find a 
reasonable solution ($n_H$ $\le$ 1600 cm$^{-3}$) by matching the 
values of $N$(C$_2$) given in Table 5.  It appeared that the values 
for $N$(C$_2$) were all too high.  Placing the present results on 
the plots of $N$(CN) versus $N$(CH), $N$(C$_2$) versus $N$(CH), and 
$N$(CN) versus $N$(C$_2$) given by Federman et al. (1994) revealed 
an interesting trend.  Our C$_2$ results for the sight lines through 
IC~443 were not distributed uniformly about the best fit, while those 
of CN were.  It appears that our derived C$_2$ column densities are 
too high by a factor of a few, but we cannot discern the cause.  As a 
result, we restricted the analysis to reproducing the CN column 
densities.  It is important to note, however, that the analysis of 
C$_2$ excitation above is not affected by this multiplicative factor 
because it relies on relative populations.

The restricted chemical analysis is able to reproduce the observed 
values for CN with gas densities of 200 to 400 cm$^{-3}$ for most 
of the sight lines (ALS~8828, HD~254346, HD~254577, HD~254755, 
HD~43703, and HD~43907).  This range in density agrees very well with 
the results of C$_2$ excitation for the directions toward HD~254346, 
HD~254577, and HD~254755.  The lone exception is the molecular gas 
toward HD~254477, where the inferred density is about a factor of 2 
higher.  This may arise because of our sight lines, only this 
direction passes within a contour for a clump (B -- see Fig. 1).  
(Since there is no CN detected toward HD~43582, this star must lie 
in front of clump H).  It seems that the material probed by our 
measurements represents the diffuse (molecular) envelope of the 
ambient cloud.

\subsection{High-velocity Gas}

We arbitrarily define ``high-velocity gas'' as deviating from the local
velocity of $-$5 km s$^{-1}$ for IC~443 by more than 25 km s$^{-1}$.  
Galactic rotation should not contribute high velocity components 
since the Galactic longitude of IC 443 is 189$^{\circ}$.  
High-velocity gas is seen mostly in Ca~{\small II} and predominantly 
with negative velocities, indicating expansion from the center of the 
system toward the Sun.  Three stars show high velocity Ca~{\small I} 
and two stars show it in the K~{\small I} line, which is rather unusual.  
High velocity CH$^+$ is seen toward HD~254477 and HD~43703.  No
other molecular lines reveal high-velocity absorption.  The only star
to show high positive velocity gas is HD~43582.  The 
$N$(Ca~{\small I})/$N$(Ca~{\small II}) ratio and the presence of the 
CH$^+$ molecule indicate that the high-velocity gas is associated with 
low densities.

Some comparison with 21 cm emission is possible, though the beam size, even
with an interferometer, was 21 $\times$ 56 arcseconds in the survey of 
Braun \& Strom (1986).  In at least one case it is possible to compare 21 
cm emission and optical absorption data.  As shown in their Fig. 7, they 
detect emission with velocities up to $-$105 km s$^{-1}$ at 
$\alpha$(1950) $=$ 6$^h$ 14$^m$ 16$^s$, $\delta$(1950) $=$ 22$^{\circ}$ 
30$^{\prime}$ 14$^{\prime\prime}$, which is fairly close to HD~254477 
(see our Fig. 1 in 1950 coordinates).  That star shows Ca~{\small II} 
absorption components at $-$64.5, $-$77.6, $-$91.4, and $-$115.7 km 
s$^{-1}$.  We see CH$^+$ absorption from the $-$64.5 and $-$77.6 km 
s$^{-1}$ components, considering the precision of our velocity scale, 
but not at the more extreme velocities.  In their Fig. 8, Braun \& Strom
show velocities from $-$97 to $-$105 km s$^{-1}$ over an arch-like 
structure in their field centered on $\alpha$ $=$ 6$^h$ 14$^m$ 15$^s$, 
$\delta$(1950) $=$ 22$^{\circ}$ 27$^{\prime}$ that is comparable to our 
high velocity Ca~{\small II} absorption.  The range in Ca~{\small II} 
velocities seen toward HD~254477, HD~254577, and HD~43582 suggests a 
shock velocity of about 100 km s$^{-1}$ for the expanding SNR.  This 
would correspond to the fast J-type shock in the analysis of 
Snell et al. (2005).

The only really high velocity CH$^+$ absorption was seen toward HD~254477, 
where high velocity Ca~{\small I} and K~{\small I} features also appeared 
at comparable velocities.  High velocity Ca~{\small I} was seen toward 
two other stars (HD~254577 and HD~43582), and K~{\small I} toward only 
one other object (HD~254577).  These three directions are most closely 
associated with the clumps seen in molecular gas, but with 
the absorption probing much lower densities.  The next highest 
velocity cloud with CH$^+$ was seen in the direction of HD~43703 at $-$30 
km s$^{-1}$.  We found no detectable CH at high velocities.  The presence 
of high velocity CH$^+$ without corresponding CH at the same velocity might 
provide a clue to the origin of interstellar CH$^+$, which remains 
mysterious despite its known presence for decades.  In the Vela Remnant, 
high velocity CH$^+$ was seen in the background star HD~72088 without 
a corresponding component in CH (Wallerstein \& Gilroy 1992).

\section{Conclusions}

Our measurements of atomic and molecular absorption toward stars 
lying behind IC~443 reveal connections to the gas studied via 
H~{\small I} and CO emission.  The molecular rich diffuse gas has a 
velocity that associates it with the ambient molecular cloud.  Its 
density is considerably lower than that inferred for the molecular 
cloud, but this is not unexpected because absorption probes the 
diffuse envelope of the cloud.  High-velocity gas is revealed by 
Ca~{\small II} absorption, which along the eastern edge of the 
nebula spans a range from $-$120 to $+$80 km s$^{-1}$.  This range 
is similar to that seen in the shocked molecular gas and its 
associated 21 cm emission.  One difference is that individual 
velocity components in the Ca~{\small II} spectra are discerned, 
suggesting that the expanding SNR created a 100 km s$^{-1}$ shock.  
The densities derived in the analysis of 21 cm 
emission, $\sim$10 cm$^{-3}$, probably pertain to the high-velocity 
gas that we observe.  The optical results presented here 
provide another facet of the interaction between a SNR remnant 
and a molecular cloud surrounding it and will aid future modeling 
efforts.

\acknowledgments
We thank Ron Snell for permission to use a digitized version of the 
figure in Huang et al. (1986).  G.W. acknowledges an enlightening 
conversation with Bruce Draine regarding the origin of CH$^+$.  
The research by Alec Hirschauer and Tim Means was supported by 
the Kenilworth Fund of the  New York Community Trust.  We made use 
of the SIMBAD database, operated at Centre de Don\'{e}es 
Astronomiques de Strasbourg, Strasbourg, France.

\begin{deluxetable}{lcccclcc}
\tablecolumns{8}
\tablewidth{0pt}
\tabletypesize{\scriptsize}
\tablecaption{IC~443 objects}
\startdata
\hline \hline \\
Name & R.A. (2000) & Dec. (2000) & $V$ & $B-V$ & Spectral Type & $E$($B-V$) & 
$M_V$ \\ \hline
ALS~8828 & 06$^h$ 16$^m$ 13.3$^s$ & $+$22$^{\circ}$ 45$^{\prime}$ 
48$^{\prime\prime}$ & 10.90 & 0.60 & B2V & 0.82 & $-$2.50 \\
HD~254346 & 06 16 57.3 & $+$22 11 42 & 9.74 & 0.39 & B2:III: & 0.63 & 
$-$3.60 \\
HD~254477 & 06 17 26.4 & $+$22 25 38 & 9.77 & 0.59 & B8V & 0.68 & 
0.10 \\
HD~254577 & 06 17 54.3 & $+$22 24 32 & 9.19 & 0.64 & B0.5II-III & 0.92 & 
$-$4.95 \\
HD~43582 & 06 18 00.3 &	$+$22 39 29 & 8.79 & 0.32 & B0IIIn & 0.62 & 
$-$5.00 \\
HD~254700 & 06 18 25.3 & $+$22 57 34 & 9.83 & 0.33 & B5V & 0.49 & 
$-$1.00 \\
HD~254755 & 06 18 31.7 & $+$22 40 45 & 8.91 & 0.42 & O9Vp & 0.73 & 
$-$4.80 \\
HD~43703 & 06 18 39.4 &	$+$23 00 28 & 8.65 & 0.33 & B1IV:p & 0.59 & 
$-$4.10 \\
HD~43753 & 06 18 59.7 &	$+$23 00 04 & 7.89 & 0.25 & B0.5III & 0.53 & 
$-$4.70 \\
HD~43871 & 06 19 34.5 &	$+$22 49 47 & 8.38 & 0.05 & A0V & 0.05 & 
1.00 \\
HD~43907 & 06 19 45.2 &	$+$22 06 38 & 8.70 & 0.30 & B1V:p & 0.56 & 
$-$3.60 \\
\enddata 
\end{deluxetable}

\begin{deluxetable}{lccccccccccccc}
\rotate
\tablecolumns{14}
\tablewidth{0pt}
\tabletypesize{\scriptsize}
\tablecaption{Equivalent Widths toward ALS~8828, HD~254346, HD~254477, 
HD~254577, HD~43582, and HD254700}
\startdata
\hline \hline\\
Line & $\lambda$ & \multicolumn{2}{c}{ALS~8828} & 
\multicolumn{2}{c}{HD~254346} & \multicolumn{2}{c}{HD~254477} & 
\multicolumn{2}{c}{HD~254577} & \multicolumn{2}{c}{HD~43582} & 
\multicolumn{2}{c}{HD~254700} \\
 &  (\AA) & $W_{\lambda}$ & $v_{LSR}$ & $W_{\lambda}$ & $v_{LSR}$ & 
$W_{\lambda}$ & $v_{LSR}$ & $W_{\lambda}$ & $v_{LSR}$ & $W_{\lambda}$ & 
$v_{LSR}$ & $W_{\lambda}$ & $v_{LSR}$ \\
 & & (m\AA) & (km s$^{-1}$) & (m\AA) & (km s$^{-1}$) & (m\AA) & 
(km s$^{-1}$) & (m\AA) & (km s$^{-1}$) & (m\AA) & (km s$^{-1}$) & 
(m\AA) & (km s$^{-1}$) \\ \hline
C$_2$ R(0) & 8757.686 & 5.9(1.3) & $-$5.4 & 3.7(2.1) & $-$2.3 & $\ldots$ & 
$\ldots$ & 6.3(0.8) & $-$9.0 & $\ldots$ & $\ldots$ & $\ldots$ & $\ldots$ \\
C$_2$ R(2) & 8753.949 & ~2.7 & $\ldots$ & 5.8(1.7) & $-$2.7 & $\ldots$ & 
$\ldots$ & 11.6(1.2) & $-$8.1 & $\ldots$ & $\ldots$ & $\ldots$ & $\ldots$ \\
C$_2$ Q(2) & 8761.194 & 5.3(1.4) & $-$5.1 & $\ldots$ & $\ldots$ & $\ldots$ & 
$\ldots$ & 15.1(1.4) & $-$8.3 & $\ldots$ & $\ldots$ & $\ldots$ & $\ldots$ \\
C$_2$ P(2) & 8766.031 & $\ldots$ & $\ldots$ & 5.0(1.9) & $-$4.2 & $\ldots$ & 
$\ldots$ & $\ldots$ & $\ldots$ & $\ldots$ & $\ldots$ & $\ldots$ & $\ldots$ \\
C$_2$ R(4) & 8751.686 & 4.8(1.9) & $-$12.2 & 10.7(2.1) & $-$1.9 & $\ldots$ & 
$\ldots$ & 7.8(1.3)\tablenotemark{a} & $-$7.5 & $\ldots$ & $\ldots$ & 
$\ldots$ & $\ldots$ \\
C$_2$ Q(4) & 8763.751 & 7.1(1.0) & $-$6.8 & 7.4(2.2) & $-$2.9 & 6.3(2.8) & 
$+$0.1 & 16.7(1.1) & $-$8.1 & $\ldots$ & $\ldots$ & $\ldots$ & $\ldots$ \\
C$_2$ P(4) & 8773.430 & $\ldots$ & $\ldots$ & $\ldots$ & $\ldots$ & $\ldots$ & 
$\ldots$ & 4.0(1.2) & $-$6.7 & $\ldots$ & $\ldots$ & $\ldots$ & $\ldots$ \\
C$_2$ R(6) & 8750.848 & $\ldots$ & $\ldots$ & $\ldots$ & $\ldots$ & $\ldots$ & 
$\ldots$ & 4.9(1.0) & $-$6.1 & $\ldots$ & $\ldots$ & $\ldots$ & $\ldots$ \\
C$_2$ Q(6) & 8767.759 & $\ldots$ & $\ldots$ & 4.8(1.9) & $-$3.8 & $\ldots$ & 
$\ldots$ & 8.2(1.0) & $-$6.5 & $\ldots$ & $\ldots$ & $\ldots$ & $\ldots$ \\
C$_2$ P(6) & 8782.308 & $\ldots$ & $\ldots$ & $\ldots$ & $\ldots$ & $\ldots$ & 
$\ldots$ & 3.4(0.8) & $-$6.8 & $\ldots$ & $\ldots$ & $\ldots$ & $\ldots$ \\
C$_2$ Q(8) & 8773.221 & 3.4(1.4)\tablenotemark{b} & $\ldots$ & $\ldots$ & 
$\ldots$ & $\ldots$ & $\ldots$ & 6.1(1.2) & $-$6.7 & $\ldots$ & $\ldots$ & 
$\ldots$ & $\ldots$ \\
C$_2$ P(8) & 8792.648 & $\ldots$ & $\ldots$ & $\ldots$ & $\ldots$ & $\ldots$ & 
$\ldots$ & 2.5(1.0) & $-$6.4 & $\ldots$ & $\ldots$ & $\ldots$ & $\ldots$ \\
C$_2$ Q(10) & 8780.141 & $\ldots$ & $\ldots$ & $\ldots$ & $\ldots$ & $\ldots$ & 
$\ldots$ & 3.4(0.9) & $-$7.2 & $\ldots$ & $\ldots$ & $\ldots$ & $\ldots$ \\
CN R(0) & 3874.610 & 12.7(1.8) & $-$9.2 & 10.7(1.0) & $-$5.7 & 13.6(1.7) & 
$-$1.5 & 24.4(1.2) & $-$9.4 & $\ldots$ & $\ldots$ & $\ldots$ & $\ldots$ \\
CN R(1) & 3874.000 & $\ldots$ & $\ldots$ & 2.9(0.8) & $-$6.2 & 8.4(1.8) & 
$-$2.2 & 13.4(1.2) & $-$7.9 & $\ldots$ & $\ldots$ & $\ldots$ & $\ldots$ \\
CN P(1) & 3875.760 & $\ldots$ & $\ldots$ & $\ldots$ & $\ldots$ & 6.5(1.5) & 
$-$1.1 & 7.1(1.1) & $-$9.2 & $\ldots$ & $\ldots$ & $\ldots$ & $\ldots$ \\
CH $A-X$ & 4300.313 & 42.3(1.3) & $-$8.5 & 21.7(0.8) & $-$3.8 & 28.0(1.7) & 
$-$2.0 & 23.4(0.6) & $-$7.0 & 12.1(0.4) & $-$8.9 & 10.6(0.9) & $-$4.6 \\
 & $\ldots$ & $\ldots$ & $\ldots$ & 4.2(0.6) & $+$8.6 & 3.6(1.7) & $+$12.8 & 
2.2(0.6) & $+$9.7 & 3.3(0.4) & $+$5.9 & $\ldots$ & $\ldots$ \\
CH $B-X$ & 3890.217 & 11.6(2.1) & $-$9.1 & $\ldots$ & $\ldots$ & $\ldots$ & 
$\ldots$ & 3.0(0.7) & $-$7.2 & 3.2(0.5) & $-$9.5 & $\ldots$ & $\ldots$ \\
CH $B-X$ & 3886.409 & 14.3(3.1) & $-$8.5 & 9.8(2.1) & $-$0.7 & 5.6(2.2) & 
$-$1.1 & 4.4(0.8) & $-$7.3 & 2.8(0.4) & $-$9.5 & 4.3(1.6) & $-$2.1 \\
CH $B-X$ & 3878.774 & $\ldots$ & $\ldots$ & $\ldots$ & $\ldots$ & $\ldots$ & 
$\ldots$ & 2.2(0.7) & $-$4.4 & $\ldots$ & $\ldots$ & $\ldots$ & $\ldots$ \\
CH$^+$ (0-0) & 4232.548 & 22.5(1.6) & $-$7.9 & 11.4(0.7) & $-$4.9 & 4.2(1.8) & 
$-$74.5 & 2.1(0.5) & $-$19.4 & 19.9(0.4) & $-$8.8 & 15.4(0.8) & $-$4.3 \\
 & $\ldots$ & 11.0(1.6) & $+$5.9 & 15.8(0.8) & $+$9.2 & 4.3(1.8) & $-$61.1 & 
19.0(0.5) & $-$5.4 & 4.4(0.4) & $+$3.1 & 4.7(0.9) & $+$8.4 \\
 & $\ldots$ & $\ldots$ & $\ldots$ & $\ldots$ & $\ldots$ & 13.8(2.2) & 
$+$1.2 & 6.8(0.5) & $+$9.2 & $\ldots$ & $\ldots$ & $\ldots$ & $\ldots$ \\
CH$^+$ (1-0) & 3957.692 & 12.9(2.5) & $-$8.6 & 6.4(1.2) & $-$3.8 & 5.7(2.6) & 
$-$1.6 & 10.4(0.9) & $-$6.8 & 8.9(0.6) & $-$8.2 & 6.9(1.2) & $-$4.3 \\
 & $\ldots$ & $\ldots$ & $\ldots$ & 6.6(1.1) & $+$9.0 & $\ldots$ & $\ldots$ & 
3.7(0.9) & $+$14.2 & 2.2(0.6) & $+$3.1 & 2.2(1.2) & $+$8.4 \\
Ca {\tiny II} K & 3933.663 & 138(3) & $-$7.9 & 12.8(1.6) & $-$75.0 & 
39.9(2.3) & $-$118.1 & 78.4(1.0) & $-$110.8 & 16.6(0.8) & $-$107.8 & 
14.3(1.6) & $-$24.9 \\
 & $\ldots$ & 131(3) & $+$5.6 & 236(2) & $-$2.2 & 154(2) & $-$91.4 & 184(1) & 
$-$58.7 & 45.8(0.8) & $-$91.5 & 115(1) & $-$5.0 \\
 & $\ldots$ & $\ldots$ & $\ldots$ & 87.2(1.8) & $+$12.3 & 133(2) & $-$77.6 & 
183(1) & $-$41.9 & 106(1) & $-$71.8 & 104(1) & $+$7.1 \\
 & $\ldots$ & $\ldots$ & $\ldots$ & $\ldots$ & $\ldots$ & 144(2) & 
$-$64.5 & 179(1) & $-$23.7 & 164(1) & $-$59.7 & 36.7(1.3) & $+$15.3 \\
 & $\ldots$ & $\ldots$ & $\ldots$ & $\ldots$ & $\ldots$ & 142(2) & 
$-$50.6 & 184(1) & $-$7.1 & 176(1) & $-$42.9 & $\ldots$ & $\ldots$ \\
 & $\ldots$ & $\ldots$ & $\ldots$ & $\ldots$ & $\ldots$ & 97.4(2.3) & 
$-$38.4 & 189(1) & $+$10.3 & 156(1) & $-$26.8 & $\ldots$ & $\ldots$ \\
 & $\ldots$ & $\ldots$ & $\ldots$ & $\ldots$ & $\ldots$ & 161(2) & 
$-$20.3 & $\ldots$ & $\ldots$ & 157(1) & $-$8.9 & $\ldots$ & $\ldots$ \\
 & $\ldots$ & $\ldots$ & $\ldots$ & $\ldots$ & $\ldots$ & 153(2) & 
$-$3.4 & $\ldots$ & $\ldots$ & 152(1) & $+$6.2 & $\ldots$ & $\ldots$ \\
 & $\ldots$ & $\ldots$ & $\ldots$ & $\ldots$ & $\ldots$ & 151(2) & 
$+$11.4 & $\ldots$ & $\ldots$ & 62.6(0.5) & $+$36.4 & $\ldots$ & $\ldots$ \\
 & $\ldots$ & $\ldots$ & $\ldots$ & $\ldots$ & $\ldots$ & $\ldots$ & 
$\ldots$ & $\ldots$ & $\ldots$ & 8.0(0.5) & $+$79.5 & $\ldots$ & $\ldots$ \\
\hline \hline\\
Line & $\lambda$ & \multicolumn{2}{c}{ALS~8828} &
\multicolumn{2}{c}{HD~254346} & \multicolumn{2}{c}{HD~254477} &
\multicolumn{2}{c}{HD~254577} & \multicolumn{2}{c}{HD~43582} &
\multicolumn{2}{c}{HD~254700} \\
 &  (\AA) & $W_{\lambda}$ & $v_{LSR}$ & $W_{\lambda}$ & $v_{LSR}$ &
$W_{\lambda}$ & $v_{LSR}$ & $W_{\lambda}$ & $v_{LSR}$ & $W_{\lambda}$ &
$v_{LSR}$ & $W_{\lambda}$ & $v_{LSR}$ \\
 & & (m\AA) & (km s$^{-1}$) & (m\AA) & (km s$^{-1}$) & (m\AA) &
(km s$^{-1}$) & (m\AA) & (km s$^{-1}$) & (m\AA) & (km s$^{-1}$) &
(m\AA) & (km s$^{-1}$) \\ \hline
Ca {\tiny II} H & 3968.468 & 116(3) & $-$7.9 & 4.7(1.3) & $-$74.0 & 8.6(2.4) & 
$-$115.7 & 43.3(0.7) & $-$110.9 & 5.7(0.6) & $-$107.2 & 4.5(2.1) & $-$29.5 \\
 & $\ldots$ & 109(3) & $+$4.8 & 171(2) & $-$2.5 & 120(3) & $-$90.4 & 164(1) & 
$-$57.3 & 23.0(0.6) & $-$90.9 & 82.2(1.4) & $-$4.3 \\
 & $\ldots$ & $\ldots$ & $\ldots$ & 65.6(1.5) & $+$9.8 & 130(3) & $-$77.7 & 
164(1) & $-$41.8 & 81.5(0.6) & $-$70.2 & 72.5(1.4) & $+$7.8 \\
 & $\ldots$ & $\ldots$ & $\ldots$ & $\ldots$ & $\ldots$ & 130(3) & 
$-$64.8 & 160(1) & $-$21.7 & 129(1) & $-$58.4 & 14.5(1.4) & $+$16.4 \\
 & $\ldots$ & $\ldots$ & $\ldots$ & $\ldots$ & $\ldots$ & 117(3) & 
$-$50.7 & 159(1) & $-$6.4 & 150(1) & $-$41.5 & $\ldots$ & $\ldots$ \\
 & $\ldots$ & $\ldots$ & $\ldots$ & $\ldots$ & $\ldots$ & 59.3(3.4) & 
$-$38.1 & 154(1) & $+$9.8 & 108(1) & $-$25.6 & $\ldots$ & $\ldots$ \\
 & $\ldots$ & $\ldots$ & $\ldots$ & $\ldots$ & $\ldots$ & 108(3) & 
$-$20.1 & $\ldots$ & $\ldots$ & 125(1) & $-$7.7 & $\ldots$ & $\ldots$ \\
 & $\ldots$ & $\ldots$ & $\ldots$ & $\ldots$ & $\ldots$ & 121(3) & 
$-$3.0 & $\ldots$ & $\ldots$ & 91.3(0.6) & $+$7.1 & $\ldots$ & $\ldots$ \\
 & $\ldots$ & $\ldots$ & $\ldots$ & $\ldots$ & $\ldots$ & 105(3) & 
$+$10.4 & $\ldots$ & $\ldots$ & 37.9(0.4) & $+$36.4 & $\ldots$ & $\ldots$ \\
 & $\ldots$ & $\ldots$ & $\ldots$ & $\ldots$ & $\ldots$ & $\ldots$ & 
$\ldots$ & $\ldots$ & $\ldots$ & 5.5(0.5) & $+$79.8 & $\ldots$ & $\ldots$ \\
Ca {\tiny I} & 4226.728 & 4.9(1.6) & $-$14.1 & 8.8(0.7) & $-$1.2 & 51.4(3.1) & 
$-$72.6 & 13.3(0.8) & $-$55.8 & 6.2(0.7) & $-$63.6 & 8.7(1.2) & $-$4.7 \\
 & $\ldots$ & 11.0(1.7) & $-$1.3 & 2.5(0.7) & $+$11.3 & 8.0(3.1) & $-$59.2 & 
18.1(0.8) & $-$46.7 & 12.8(0.7) & $-$40.8 & 9.2(1.2) & $+$7.6 \\
 & $\ldots$ & $\ldots$ & $\ldots$ & $\ldots$ & $\ldots$ & 11.9(2.3) & 
$-$1.1 & 8.7(0.6) & $-$21.8 & 8.0(0.7) & $-$10.2 & $\ldots$ & $\ldots$ \\
 & $\ldots$ & $\ldots$ & $\ldots$ & $\ldots$ & $\ldots$ & $\ldots$ & 
$\ldots$ & 19.9(0.6) & $-$7.5 & 2.2(0.7) & $+$3.6 & $\ldots$ & $\ldots$ \\
 & $\ldots$ & $\ldots$ & $\ldots$ & $\ldots$ & $\ldots$ & $\ldots$ & 
$\ldots$ & 6.9(0.6) & $+$10.3 & $\ldots$ & $\ldots$ & $\ldots$ & $\ldots$ \\
K {\tiny I} & 7698.974 & 216(2) & $-$8.1 & 176(1) & $-$1.4 & 9.5(2.3) & 
$-$82.7 & 7.9(1.1) & $-$55.4 & 127(2) & $-$9.1 & 134(2) & $-$5.2 \\
 & $\ldots$ & 51.8(1.9) & $+$5.7 & 51.1(1.2) & $+$9.9 & 45.4(2.3) & $-$65.0 & 
14.4(1.1) & $-$46.9 & 25.0(1.6) & $+$5.5 & 30.3(1.5) & $+$6.9 \\
 & $\ldots$ & $\ldots$ & $\ldots$ & $\ldots$ & $\ldots$ & 12.0(2.3) & 
$-$52.6 & 7.0(1.2) & $-$20.1 & $\ldots$ & $\ldots$ & 9.3(1.5) & $+$15.7 \\
 & $\ldots$ & $\ldots$ & $\ldots$ & $\ldots$ & $\ldots$ & 170(2) & $-$1.5 & 
221(1) & $-$5.9 & $\ldots$ & $\ldots$ & $\ldots$ & $\ldots$ \\
 & $\ldots$ & $\ldots$ & $\ldots$ & $\ldots$ & $\ldots$ & 23.9(2.2) & 
$+$10.0 & 33.3(1.2) & $+$8.6 & $\ldots$ & $\ldots$ & $\ldots$ & $\ldots$ \\
\enddata
\tablenotetext{a}{May include absorption from R(8).}
\tablenotetext{b}{May include absorption from P(4).}
\end{deluxetable}

\begin{deluxetable}{lccccccccccc}
\rotate
\tablecolumns{12}
\tablewidth{0pt}
\tabletypesize{\scriptsize}
\tablecaption{Equivalent Widths toward HD~254755, HD~43703, HD~43753,
HD~43871, and HD~43907}
\startdata
\hline \hline\\
Line & $\lambda$ & \multicolumn{2}{c}{HD~254755} &
\multicolumn{2}{c}{HD~43703} & \multicolumn{2}{c}{HD~43753} &
\multicolumn{2}{c}{HD~43871} & \multicolumn{2}{c}{HD~43907} \\
 &  (\AA) & $W_{\lambda}$ & $v_{LSR}$ & $W_{\lambda}$ & $v_{LSR}$ &
$W_{\lambda}$ & $v_{LSR}$ & $W_{\lambda}$ & $v_{LSR}$ & $W_{\lambda}$ &
$v_{LSR}$ \\
 & & (m\AA) & (km s$^{-1}$) & (m\AA) & (km s$^{-1}$) & (m\AA) &
(km s$^{-1}$) & (m\AA) & (km s$^{-1}$) & (m\AA) & (km s$^{-1}$) \\ \hline
C$_2$ R(0) & 8757.686 & 2.5(0.7) & $-$6.7 & $\ldots$ & $\ldots$ & 
$\ldots$ & $\ldots$ & $\ldots$ & $\ldots$ & $\ldots$ & $\ldots$ \\
C$_2$ R(2) & 8753.949 & 5.6(0.9) & $-$3.8 & $\ldots$ & $\ldots$ & 
$\ldots$ & $\ldots$ & $\ldots$ & $\ldots$ & $\ldots$ & $\ldots$ \\
C$_2$ Q(2) & 8761.194 & 5.3(0.9) & $-$4.7 & $\ldots$ & $\ldots$ & 
$\ldots$ & $\ldots$ & $\ldots$ & $\ldots$ & $\ldots$ & $\ldots$ \\
C$_2$ P(2) & 8766.031 & 1.9(0.8) & $-$2.7 & $\ldots$ & $\ldots$ 
& $\ldots$ & $\ldots$ & $\ldots$ & $\ldots$ & $\ldots$ & $\ldots$ \\
C$_2$ R(4) & 8751.686 & 3.5(0.9) & $-$6.8 & $\ldots$ & $\ldots$ & 
$\ldots$ & $\ldots$ & $\ldots$ & $\ldots$ & $\ldots$ & $\ldots$ \\
C$_2$ Q(4) & 8763.751 & 10.2(1.2) & $-$5.7 & 7.5(1.3) & $-$3.3 & 
$\ldots$ & $\ldots$ & $\ldots$ & $\ldots$ & 5.4(1.3) & $-$5.4 \\
C$_2$ P(4) & 8773.430 & 2.6(0.9) & $-$5.6 & $\ldots$ & $\ldots$ & 
$\ldots$ & $\ldots$ & $\ldots$ & $\ldots$ & $\ldots$ & $\ldots$ \\
C$_2$ Q(6) & 8767.759 & 2.8(0.8) & $-$5.3 & $\ldots$ & $\ldots$ & 
$\ldots$ & $\ldots$ & $\ldots$ & $\ldots$ & $\ldots$ & $\ldots$ \\
C$_2$ Q(8) & 8773.221 & 3.1(0.9) & $-$4.8 & $\ldots$ & $\ldots$ & 
$\ldots$ & $\ldots$ & $\ldots$ & $\ldots$ & $\ldots$ & $\ldots$ \\
C$_2$ Q(10) & 8780.141 & 3.2(1.0) & $-$8.1 & $\ldots$ & $\ldots$ & 
$\ldots$ & $\ldots$ & $\ldots$ & $\ldots$ & $\ldots$ & $\ldots$ \\
CN R(0) & 3874.610 & 7.8(0.6) & $-$6.3 & 2.2(0.4) & $-$5.7 & $\ldots$ &
$\ldots$ & $\ldots$ & $\ldots$ & 2.0(0.6) & $-$12.2 \\
CN R(1) & 3874.000 & 4.2(0.7) & $-$7.0 & 1.6(0.5) & $-$10.1 & $\ldots$ &
$\ldots$ & $\ldots$ & $\ldots$ & $\ldots$ & $\ldots$ \\
CH $A-X$ & 4300.313 & 25.6(0.5) & $-$6.0 & 20.0(0.6) & $-$5.8 & 10.4(0.7) &
$-$5.4 & $\ldots$ & $\ldots$ & 10.9(1.1) & $-$10.0 \\
 & $\ldots$ & 2.7(0.5) & $+$5.0 & $\ldots$ & $\ldots$ & $\ldots$ & $\ldots$ &
$\ldots$ & $\ldots$ & 5.1(1.1) & $+$2.4 \\
CH $B-X$ & 3890.217 & 5.1(0.6) & $-$6.7 & 2.8(0.6) & $-$5.1 & $\ldots$ & 
$\ldots$ & $\ldots$ & $\ldots$ & $\ldots$ & $\ldots$ \\
CH $B-X$ & 3886.409 & 7.2(0.6) & $-$3.9 & 4.0(0.6) & $-$6.3 & $\ldots$ & 
$\ldots$ & $\ldots$ & $\ldots$ & 3.2(0.8) & $-$11.6 \\
 & $\ldots$ & 1.1(0.6) & $+$8.9 & $\ldots$ & $\ldots$ & $\ldots$ & $\ldots$ & 
$\ldots$ & $\ldots$ & 1.5(0.8) & $-$1.8 \\
CH$^+$ (0-0) & 4232.548 & 25.1(0.5) & $-$6.5 & 2.4(0.5) & $-$30.8 & 
3.4(1.8) & $-$20.4 & $\ldots$ & $\ldots$ & 11.4(0.8) & $-$7.8 \\
 & $\ldots$ & 3.8(0.5) & $+$8.7 & 27.2(0.6) & $-$6.3 & 16.1(1.4) & $-$5.6 & 
$\ldots$ & $\ldots$ & 8.0(0.8) & $+$0.4 \\
 & $\ldots$ & $\ldots$ & $\ldots$ & 4.3(0.7) & $+$6.9 & $\ldots$ & $\ldots$ & 
$\ldots$ & $\ldots$ & $\ldots$ & $\ldots$ \\
CH$^+$ (1-0) & 3957.692 & 15.2(0.5) & $-$6.4 & 12.3(0.7) & $-$6.5 & 
9.4(1.5) & $-$7.0 & $\ldots$ & $\ldots$ & 10.2(1.3) & $-$4.7 \\
 & $\ldots$ & 2.7(0.5) & $+$7.2 & 2.2(0.6) & $+$3.1 & $\ldots$ & $\ldots$ & 
$\ldots$ & $\ldots$ & 4.3(1.3) & $+$0.2 \\
Ca {\tiny II} K & 3933.663 & 15.2(1.0) & $-$74.7 & 4.1(1.0) & $-$36.2 & 
122(2) & $-$15.1 & 92.3(0.8) & $-$4.2 & 20.0(1.0) & $-$78.8 \\
 & $\ldots$ & 47.6(1.3) & $-$24.9 & 152(1) & $-$6.0 & 114(2) & $-$1.6 & 
8.2(0.5) & $+$5.8 & 5.2(0.9) & $-$64.3 \\
 & $\ldots$ & 152(1) & $-$7.3 & 134(1) & $+$8.1 & 108(2) & $+$8.7 & $\ldots$ & 
$\ldots$ & 131(1) & $-$10.2 \\
 & $\ldots$ & 151(1) & $+$7.3 & $\ldots$ & $\ldots$ & $\ldots$ & $\ldots$ & 
$\ldots$ & $\ldots$ & 140(1) & $+$2.6 \\
Ca {\tiny II} H & 3968.468 & 7.8(0.9) & $-$77.2 & 6.4(1.0) & $-$31.9 & 
68.0(1.5) & $-$15.6 & 41.9(0.9) & $-$4.3 & 10.3(1.1) & $-$78.9 \\
 & $\ldots$ & 25.4(1.1) & $-$24.6 & 120(1) & $-$5.6 & 105(2) & $-$2.3 & 
4.6(0.8) & $+$6.9 & 3.0(0.7) & $-$63.3 \\
 & $\ldots$ & 124(1) & $-$5.5 & 95.8(1.0) & $+$7.9 & 63.6(1.5) & $+$8.6 & 
$\ldots$ & $\ldots$ & 91.5(1.3) & $-$10.0 \\
 & $\ldots$ & 90.0(1.1) & $+$7.4 & $\ldots$ & $\ldots$ & $\ldots$ & $\ldots$ & 
$\ldots$ & $\ldots$ & 117(1) & $+$1.7 \\
Ca {\tiny I} & 4226.728 & 12.4(0.7) & $-$5.5 & 10.6(0.6) & $-$6.3 & 
7.5(1.0) & $-$19.1 & $\ldots$ & $\ldots$ & 6.0(0.8) & $-$9.0 \\
 & $\ldots$ & $\ldots$ & $\ldots$ & 5.4(0.6) & $+$6.9 & 6.6(1.0) & $-$1.2 & 
$\ldots$ & $\ldots$ & 9.2(0.8) & $+$2.9 \\
K {\tiny I} & 7698.974 & 158(1) & $-$5.0 & 194(1) & $-$5.5 & 14.3(1.3) & 
$-$18.1 & 2.7(0.8) & $+$5.9 & 107(1) & $-$10.7 \\
 & $\ldots$ & 42.9(0.8) & $+$9.3 & 25.2(1.0) & $+$6.8 & 118(1) & $-$3.9 & 
$\ldots$ & $\ldots$ & 68.5(1.0) & $-$0.2 \\
 & $\ldots$ & $\ldots$ & $\ldots$ & $\ldots$ & $\ldots$ & 27.9(1.3) & 
$+$5.4 & $\ldots$ & $\ldots$ & $\ldots$ & $\ldots$ \\
\enddata
\end{deluxetable}

\begin{deluxetable}{cccccccc}
\tablecolumns{8}
\tablewidth{0pt}
\tabletypesize{\scriptsize}
\tablecaption{C$_2$ Results}
\startdata
\hline \hline\\
$J$ & \multicolumn{7}{c}{$N$($J$) (10$^{12}$ cm$^{-2}$)} \\ \cline{2-8}
 & ALS~8828 & HD~254346 & HD~254477 & HD~254577 & HD~254755 & HD~43703 & 
HD~43907 \\ \hline
0 & $8.9 \pm 2.0$ & $5.5 \pm 3.2$ & $\ldots$ & $9.5 \pm 1.2$ & $3.7 \pm 1.1$ & 
$\ldots$ & $\ldots$ \\
2 & $14 \pm 3$ & $25 \pm 7$ & $\ldots$ & $46 \pm 4$ & $19 \pm 2$ & 
$\ldots$ & $\ldots$ \\
4 & $21 \pm 4$ & $32 \pm 6$ & $19 \pm 9$ & $46 \pm 3$ & $23 \pm 3$ & 
$23 \pm 4$ & $16 \pm 4$ \\
6 & $\ldots$ & $14 \pm 6$ & $\ldots$ & $25 \pm 2$ & $8.3 \pm 2.4$ & 
$\ldots$ & $\ldots$ \\
8 & $\ldots$ & $\ldots$ & $\ldots$ & $18 \pm 3$ & $9.2 \pm 2.7$ & 
$\ldots$ & $\ldots$ \\
10 & $\ldots$ & $\ldots$ & $\ldots$ & $10 \pm 3$ & $9.5 \pm 3.0$ & 
$\ldots$ & $\ldots$ \\
\enddata
\end{deluxetable}

\begin{deluxetable}{cccccccc}
\tablecolumns{8}
\tablewidth{0pt}
\tabletypesize{\scriptsize}
\tablecaption{Component Structure}
\startdata
\hline \hline\\
$v_{LSR}$ & $N$(Ca {\tiny II}) & $N$(Ca {\tiny I}) & $N$(K {\tiny I}) & 
$N$(CH$^+$) & $N$(CH) & $N$(C$_2$) & $N$(CN) \\
km s$^{-1}$ & 10$^{11}$ cm$^{-2}$ & 10$^{9}$ cm$^{-2}$ & 10$^{11}$ cm$^{-2}$ & 
10$^{12}$ cm$^{-2}$ & 10$^{12}$ cm$^{-2}$ & 10$^{12}$ cm$^{-2}$ & 
10$^{12}$ cm$^{-2}$ \\ \hline
\multicolumn{8}{c}{ALS~8828} \\ \hline
$-$7.9 & $52-120$ & $18 \pm 7$ & $130 \pm 10$ & $30 \pm 3$ & $72 \pm 3$ & 
$66 \pm 14$ & $4.6 \pm 0.8$ \\
$+$5.6 & $45-100$ & $43 \pm 8$ & $3.6 \pm 0.2$ & $13 \pm 3$ & $\ldots$ & 
$\ldots$ & $\ldots$ \\ \hline
\multicolumn{8}{c}{HD~254346} \\ \hline
$-$75.0 & $1.4 \pm 0.2$ & $\ldots$ & $\ldots$ & $\ldots$ & $\ldots$ & 
$\ldots$ & $\ldots$ \\
$-$2.2 & $52-59$ & $34 \pm 3$ & $41 \pm 1$ & $14 \pm 1$ & $30 \pm 2$ & 
$94 \pm 25$ & $3.5 \pm 0.5$ \\
 & $4.6 \pm 0.8$ \\
$+$12.3 & $20-28$ & $9.2 \pm 2.6$ & $3.5 \pm 0.1$ & $20 \pm 1$ & $5.2 \pm 0.8$ & 
$\ldots$ & $\ldots$ \\ \hline
\multicolumn{8}{c}{HD~254477} \\ \hline
$-$118.1 & $4.1 \pm 0.3$ & $\ldots$ & $\ldots$ & $\ldots$ & $\ldots$ & 
$\ldots$ & $\ldots$ \\
$-$91.4 & $42-59$ & $\ldots$ & $\ldots$ & $\ldots$ & $\ldots$ & 
$\ldots$ & $\ldots$ \\
$-$77.6 & $>980$ & $290 \pm 30$ & $0.55 \pm 0.14$ & $4.9 \pm 2.2$ & 
$\ldots$ & $\ldots$ & $\ldots$ \\
$-$64.5 & $100-700$ & $30 \pm 13$ & $3.1 \pm 0.2$ & $5.0 \pm 2.2$ & 
$\ldots$ & $\ldots$ & $\ldots$ \\
$-$50.6 & $49-83$ & $\ldots$ & $0.70 \pm 0.15$ & $\ldots$ & $\ldots$ & 
$\ldots$ & $\ldots$ \\
$-$38.4 & $12-17$ & $\ldots$ & $\ldots$ & $\ldots$ & $\ldots$ & 
$\ldots$ & $\ldots$ \\
$-$20.3 & $28-34$ & $\ldots$ & $\ldots$ & $\ldots$ & $\ldots$ & 
$\ldots$ & $\ldots$ \\
$-$3.4 & $44-64$ & $47 \pm 10$ & $36 \pm 2$ & $16 \pm 3$ & $40 \pm 3$ & 
$63 \pm 30$ & $6.7 \pm 1.1$ \\
$+$11.4 & $29-35$ & $\ldots$ & $1.5 \pm 0.2$ & $\ldots$ & $4.4 \pm 2.2$ & 
$\ldots$ & $\ldots$ \\ \hline
\multicolumn{8}{c}{HD~254577} \\ \hline
$-$110.8 & $8.9-10.0$ & $\ldots$ & $\ldots$ & $\ldots$ & $\ldots$ & 
$\ldots$ & $\ldots$ \\
$-$58.7 & $170-230$ & $53 \pm 4$ & $0.46 \pm 0.07$ & $\ldots$ & 
$\ldots$ & $\ldots$ & $\ldots$ \\
$-$41.9 & $170-280$ & $74 \pm 4$ & $0.85 \pm 0.07$ & $\ldots$ & 
$\ldots$ & $\ldots$ & $\ldots$ \\
$-$23.7 & $160-270$ & $33 \pm 3$ & $0.40 \pm 0.07$ & $2.4 \pm 0.6$ & 
$\ldots$ & $\ldots$ & $\ldots$ \\
$-$7.1 & $110-140$ & $83 \pm 3$ & $150 \pm 10$ & $24 \pm 1$ & $32 \pm 1$ & 
$154 \pm 17$ & $11 \pm 1$ \\
$+$10.3 & $73-83$ & $26 \pm 2$ & $2.1 \pm 0.1$ & $8.1 \pm 0.6$ & 
$2.7 \pm 0.8$ & $\ldots$ & $\ldots$ \\ \hline
\multicolumn{8}{c}{HD~43582} \\ \hline
$-$107.8 & $1.6 \pm 0.1$ & $\ldots$ & $\ldots$ & $\ldots$ & $\ldots$ & 
$\ldots$ & $\ldots$ \\
$-$91.5 & $5.2 \pm 0.1$ & $\ldots$ & $\ldots$ & $\ldots$ & $\ldots$ & 
$\ldots$ & $\ldots$ \\
$-$71.8 & $29-33$ & $\ldots$ & $\ldots$ & $\ldots$ & $\ldots$ & 
$\ldots$ & $\ldots$ \\
$-$59.7 & $52-88$ & $23 \pm 3$ & $\ldots$ & $\ldots$ & $\ldots$ & 
$\ldots$ & $\ldots$ \\
$-$42.9 & $86-110$ & $50 \pm 3$ & $\ldots$ & $\ldots$ & $\ldots$ & 
$\ldots$ & $\ldots$ \\
$-$26.8 & $31-33$ & $\ldots$ & $\ldots$ & $\ldots$ & $\ldots$ & 
$\ldots$ & $\ldots$ \\
$-$8.9 & $52-60$ & $30 \pm 3$ & $15 \pm 1$ & $26 \pm 1$ & $16 \pm 1$ & 
$\ldots$ & $\ldots$ \\
$+$6.2 & $22-23$ & $8.1 \pm 2.6$ & $1.5 \pm 0.1$ & $5.2 \pm 0.5$ & 
$4.0 \pm 0.5$ & $\ldots$ & $\ldots$ \\
$+$36.4 & $9.0-10.0$ & $\ldots$ & $\ldots$ & $\ldots$ & $\ldots$ & 
$\ldots$ & $\ldots$ \\
$+$79.5 & $0.99 \pm 0.05$ & $\ldots$ & $\ldots$ & $\ldots$ & $\ldots$ & 
$\ldots$ & $\ldots$ \\ \hline
\multicolumn{8}{c}{HD~254700} \\ \hline
$-$24.9 & $1.5 \pm 0.2$ & $\ldots$ & $\ldots$ & $\ldots$ & $\ldots$ & 
$\ldots$ & $\ldots$ \\
$-$5.0 & $24-28$ & $33 \pm 5$ & $17 \pm 1$ & $18 \pm 2$ & $14 \pm 2$ & 
$\ldots$ & $\ldots$ \\
$+$7.1 & $20-23$ & $35 \pm 5$ & $1.9 \pm 0.1$ & $5.4 \pm 1.0$ & $\ldots$ & 
$\ldots$ & $\ldots$ \\
$+$15.3 & $4.1 \pm 0.1$ & $\ldots$ & $0.54 \pm 0.09$ & $\ldots$ & $\ldots$ & 
$\ldots$ & $\ldots$ \\ \hline
\multicolumn{8}{c}{HD~254755} \\ \hline
$-$74.7 & $1.7 \pm 0.1$ & $\ldots$ & $\ldots$ & $\ldots$ & $\ldots$ &
$\ldots$ & $\ldots$ \\
$-$24.9 & $5.6 \pm 0.1$ & $\ldots$ & $\ldots$ & $\ldots$ & $\ldots$ &
$\ldots$ & $\ldots$ \\
$-$7.3 & $59-67$ & $49 \pm 3$ & $27 \pm 1$ & $35 \pm 1$ & $37 \pm 1$ & 
$73 \pm 15$ & $3.2 \pm 0.5$ \\
$+$7.3 & $21-22$ & $\ldots$ & $2.9 \pm 0.1$ & $4.8 \pm 0.5$ & $3.4 \pm 0.6$ & 
$\ldots$ & $\ldots$ \\ \hline
 \\
 \\
 \\
 \\
\hline \hline\\
$v_{LSR}$ & $N$(Ca {\tiny II}) & $N$(Ca {\tiny I}) & $N$(K {\tiny I}) &
$N$(CH$^+$) & $N$(CH) & $N$(C$_2$) & $N$(CN) \\
km s$^{-1}$ & 10$^{11}$ cm$^{-2}$ & 10$^{9}$ cm$^{-2}$ & 10$^{11}$ cm$^{-2}$ &
10$^{12}$ cm$^{-2}$ & 10$^{12}$ cm$^{-2}$ & 10$^{12}$ cm$^{-2}$ &
10$^{12}$ cm$^{-2}$ \\ \hline
\multicolumn{8}{c}{HD~43703} \\ \hline
$-$36.2 & $0.68 \pm 0.10$ & $\ldots$ & $\ldots$ & $2.8 \pm 0.6$ & $\ldots$ &
$\ldots$ & $\ldots$ \\
$-$6.0 & $48-55$ & $41 \pm 3$ & $67 \pm 2$ & $35 \pm 1$ & $27 \pm 1$ &
$77 \pm 13$ & $1.0 \pm 0.3$ \\
$+$8.1 & $29-32$ & $20 \pm 3$ & $1.6 \pm 0.1$ & $5.0 \pm 0.8$ & $\ldots$ &
$\ldots$ & $\ldots$ \\ \hline
\multicolumn{8}{c}{HD~43753} \\ \hline
$-$15.1 & $14-16$ & $28 \pm 4$ & $0.85 \pm 0.05$ & $4.0 \pm 2.1$ & $\ldots$ & 
$\ldots$ & $\ldots$ \\
$-$1.6 & $>$100 & $25 \pm 4$ & $13 \pm 1$ & $20 \pm 2$ & $13 \pm 1$ & 
$\ldots$ & $\ldots$ \\
$+$8.7 & $14-16$ & $\ldots$ & $1.7 \pm 0.1$ & $\ldots$ & $\ldots$ & 
$\ldots$ & $\ldots$ \\ \hline
\multicolumn{8}{c}{HD~43871} \\ \hline
$-$4.2 & $9.6 \pm 0.2$ & $\ldots$ & $0.15 \pm 0.05$ & $\ldots$ & $\ldots$ & 
$\ldots$ & $\ldots$ \\
$+$5.8 & $0.94 \pm 0.06$ & $\ldots$ & $\ldots$ & $\ldots$ & $\ldots$ & 
$\ldots$ & $\ldots$ \\ \hline
\multicolumn{8}{c}{HD~43907} \\ \hline
$-$78.8 & $2.3 \pm 0.1$ & $\ldots$ & $\ldots$ & $\ldots$ & $\ldots$ & 
$\ldots$ & $\ldots$ \\
$-$64.3 & $0.62 \pm 0.08$ & $\ldots$ & $\ldots$ & $\ldots$ & $\ldots$ & 
$\ldots$ & $\ldots$ \\
$-$10.2 & $26-29$ & $23 \pm 4$ & $10 \pm 1$ & $15 \pm 1$ & $14 \pm 2$ & 
$53 \pm 13$ & $0.68 \pm 0.21$ \\
$+$2.6 & $59-74$ & $35 \pm 4$ & $5.2 \pm 0.1$ & $9.6 \pm 0.9$ & 
$6.4 \pm 1.3$ & $\ldots$ & $\ldots$ \\
\enddata
\end{deluxetable}

\clearpage

\begin{figure}
\begin{center}
\includegraphics[scale=0.9]{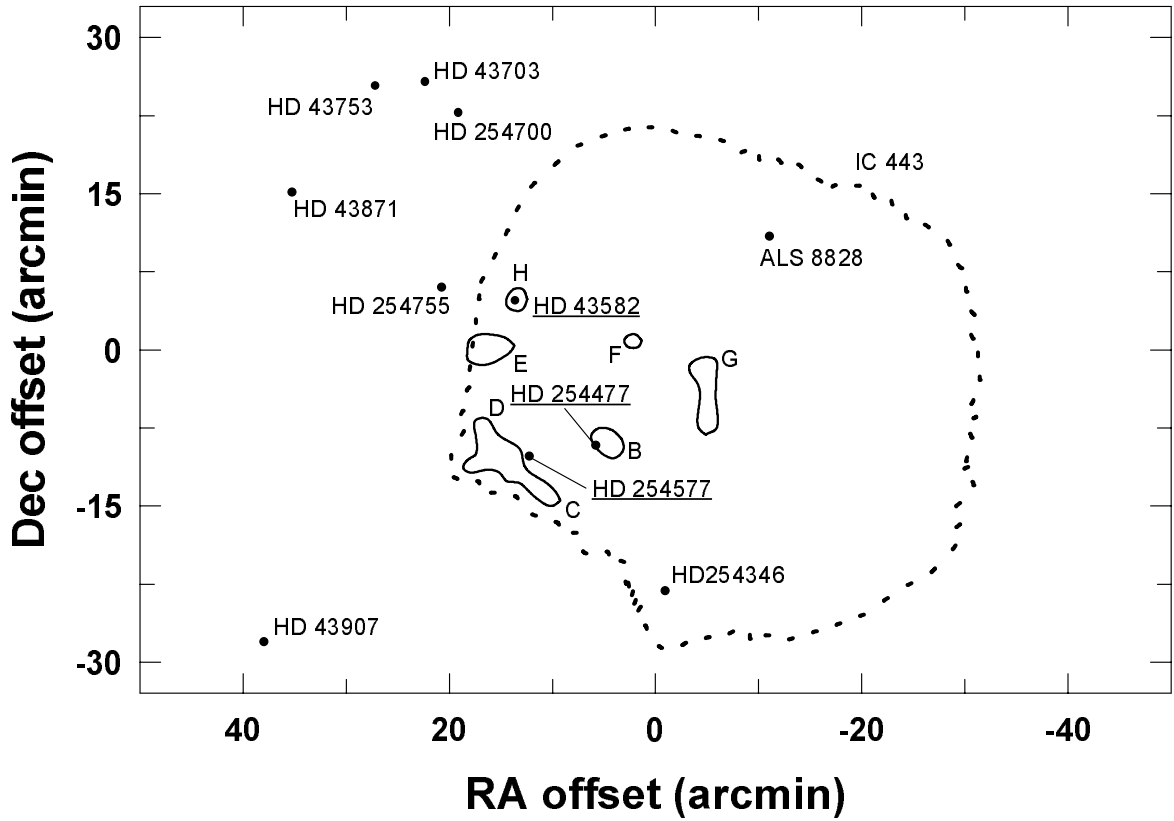}
\end{center}
\vspace{0.3in}
\caption{Map showing the relationship between IC~443, the molecular 
clumps noted by Huang et al. (1986) and shown by solid contours, and the 
directions analyzed here labeled by HD numbers.  The dashed line 
indicates the extent of the radio continuum emission.  Underlined HD 
numbers are sight lines with Ca~{\small II} absorption spread over 
100 km s$^{-1}$ (\S3.3).  The offsets are relative to RA(1950) $=$ 
6$^h$ 14$^m$ and DEC(1950) $=$ 22$^{\circ}$ 36$^{\prime}$.}
\end{figure}

\clearpage

\begin{figure}
\begin{center}
\includegraphics[scale=0.7]{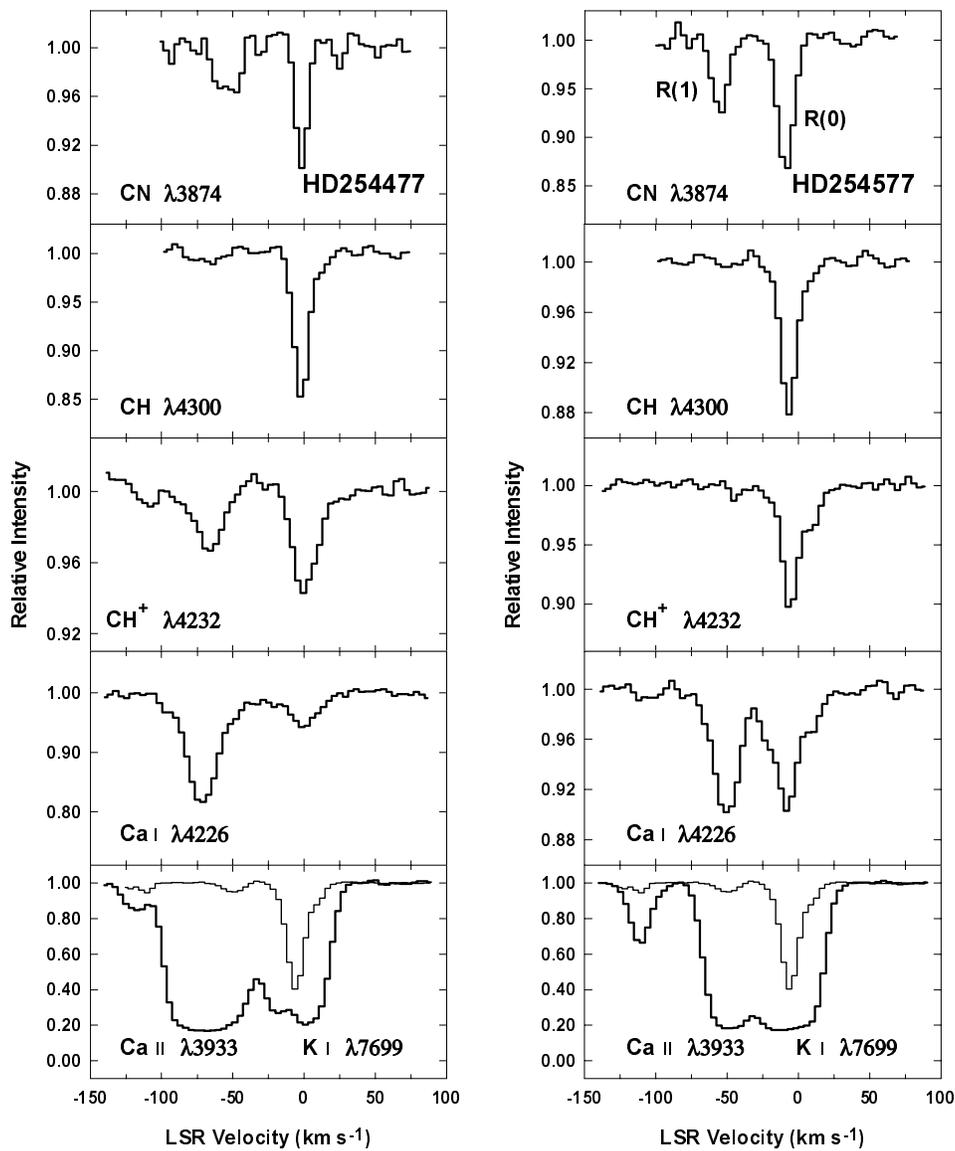}
\end{center}
\vspace{0.3in}
\caption{Spectra showing CN, CH, CH$^+$, Ca~{\small I}, Ca~{\small II}, and 
K~{\small I} absorption toward HD~254477 and HD~254577.  K~{\small I} 
$\lambda$7699 is indicated by the thin line in the bottom panel, showing 
the different velocity ranges probed by neutral and ionized gas.  The 
K~{\small I} features near $-$125 km s$^{-1}$ are telluric.  The individual 
panels have different vertical scales.}
\end{figure}

\clearpage

\begin{figure}
\begin{center}
\includegraphics[scale=0.7]{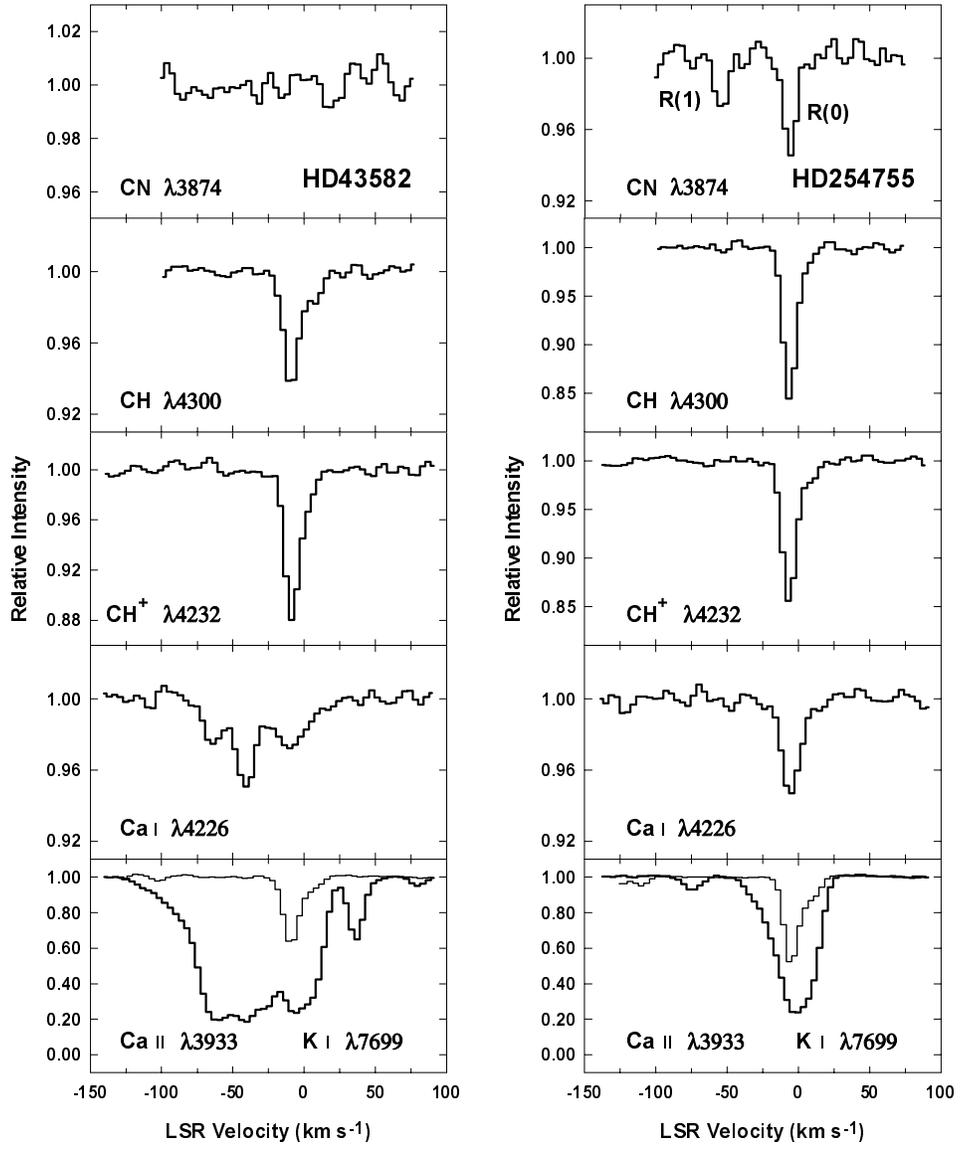}
\end{center}
\vspace{0.3in}
\caption{Same as Fig. 2 for HD~43582 and HD~254755.}
\end{figure}

\clearpage

\begin{figure}
\begin{center}
\includegraphics[scale=0.7]{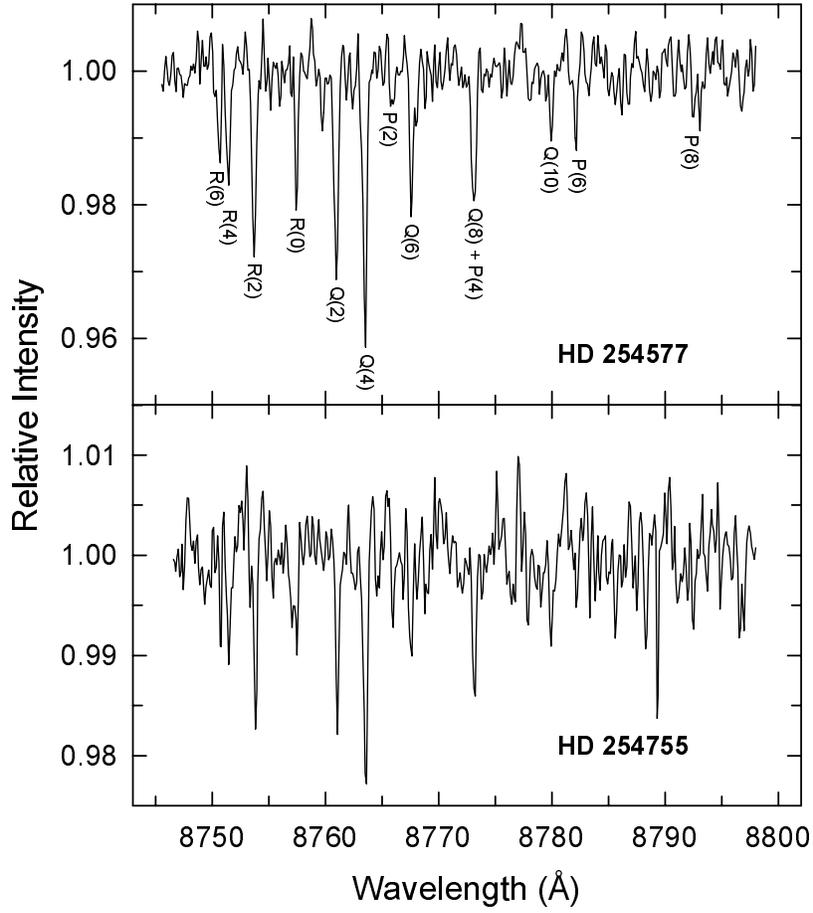}
\end{center}
\vspace{0.3in}
\caption{C$_2$ spectra toward HD~254577 and HD~254755.  The features to 
the right of the P(8) line in the top panel and near 8790 \AA\ in the 
bottom are instrumental artefacts.}
\end{figure}

\clearpage

\end{document}